\documentclass[prl, showpacs, twocolumn, reprint]{revtex4}
\usepackage{hyperref}
\usepackage[dvipdfm]{graphicx}
\usepackage{amsmath}
\usepackage{amssymb}
\usepackage{amsfonts}
\usepackage{booktabs}    
\usepackage{subfigure}
\usepackage{longtable}
\usepackage{enumerate}
\usepackage{multirow}
\usepackage{color}

\begin{document}
\title{Experimental evidence for a Bragg glass density wave phase  in a transition-metal dichalcogenide}

\author{Jun-ichi Okamoto}
\email{okamoto@phys.columbia.edu}
\affiliation{Department of Physics, Columbia University, 538 West 120th Street, New York, New York 10027, USA}

\author{Carlos J. Arguello}
\affiliation{Department of Physics, Columbia University, 538 West 120th Street, New York, New York 10027, USA}

\author{Ethan P. Rosenthal}
\affiliation{Department of Physics, Columbia University, 538 West 120th Street, New York, New York 10027, USA}

\author{Abhay N. Pasupathy}
\affiliation{Department of Physics, Columbia University, 538 West 120th Street, New York, New York 10027, USA}

\author{Andrew J. Millis}
\affiliation{Department of Physics, Columbia University, 538 West 120th Street, New York, New York 10027, USA}

\date{\today}

\begin{abstract}
Analysis of the spatial dependence of current-voltage characteristics obtained from scanning tunneling microscopy experiments indicates that the charge density wave (CDW) occurring in  NbSe$_2$  is subject to locally strong pinning by a non-negligible density of defects, but that on the length scales accessible in this experiment the material is  in a ``Bragg glass'' phase where dislocations and anti-dislocations occur in bound pairs and free dislocations are not observed. 
A Landau theory-based analysis is presented showing  how a strong local modulation may produce only a weak long range effect on the CDW  phase. 

\end{abstract}

\maketitle

The effect of disorder on the properties of condensed matter systems is important both in terms of fundamental physics and of technological applications.  In charge density wave (CDW) systems, randomly positioned impurities provide a random field which couples linearly to the order parameter \cite{Gruner1988}.  Theory dating back to the 1970s indicates that if the impurity potential is strong enough, the random field destroys the charge density wave completely, leading to a phase with exponentially decaying correlations and a correlation length of the order of the mean distance between impurities \cite{Fukuyama1978, Lee1979}. Subsequent work revised this picture, showing that in spatial dimensions $d=3$, weak impurity pinning may lead instead to a topologically ordered ``Bragg glass'' phase with power-law density correlations \cite{Feigel'man1989, Nattermann1990, Bouchaud1991, Bouchaud1992, Korshunov1993, Giamarchi1994, Giamarchi1995, Giamarchi1997, Rosso2004}.  

While the physics of random field systems has been of intense theoretical interest, experimental information has mainly come from transport and scattering measurements which average over large sample volumes \cite{Gruner1988, Sweetland1990, DeLand1991, Yaron1994, Klein2001, Ravy2006}. An important exception is the flux lattice decoration experiments which provided important early support to the Bragg glass picture for vortices in superconductors \cite{Murray1990, Grier1991}. The development of stable scanning tunneling spectroscopy (STS) techniques which provide atomic-resolution imaging of local electronic density over wide fields of view has opened up new avenues for investigation of fundamental electronic physics,  in particular providing  real-space information on the effects of disorder on electronically ordered states \cite{Dai1992, Dai1993}. In this paper, we present an analysis of scanning tunneling spectroscopy measurements carried out on NbSe$_2$, a representative charge density wave system. The analysis motivates a Landau theory which provides insights into the effects of strong pinning in charge density wave systems.  

NbSe$_2$ is a quasi-two dimensional material. Its unit cell consists of two blocks of Se-Nb-Se layers; the Nb atoms in each layer form a triangular lattice and the electrical conductivity is strongly anisotropic, being much larger for in-plane currents than for currents flowing perpendicular to the layers \cite{Dordevic2001}.  Scattering measurements \cite{Moncton1977} indicate that a second order phase transition occurs at $T_c \approx 34$ K; below this temperature a charge density wave forms. The charge density wave involves condensation of electronic density at three wavevectors ${\vec Q}_{i=1,2,3}$ related by $120^\circ$ rotations. $|{\vec Q}_i |  \approx {\vec G}_i/3\approx 0.7$ \AA$^{-1}$ with  $\vec{G}_i$ the smallest nonzero reciprocal lattice vectors. We may write the modulation of the electron density $\delta \rho$  in the charge density wave phase as 
\begin{equation}
\delta \rho(x) =  \sum_{i=1}^3  \Re \left( \psi_i (\vec{x}) e^{i\vec{Q}_i \cdot \vec{x}} \right)
\label{rho_CDW} 
\end{equation}
The CDW order parameters $\psi_i$ are complex numbers which may be written in terms of a real magnitude $\eta_i$ and a  phase $\phi_i$. Deviations from perfect charge density wave order involve spatial variations of $\eta$ and $\phi$.

\begin{figure*}[!htb]
\centering
\includegraphics[width=2.0\columnwidth]{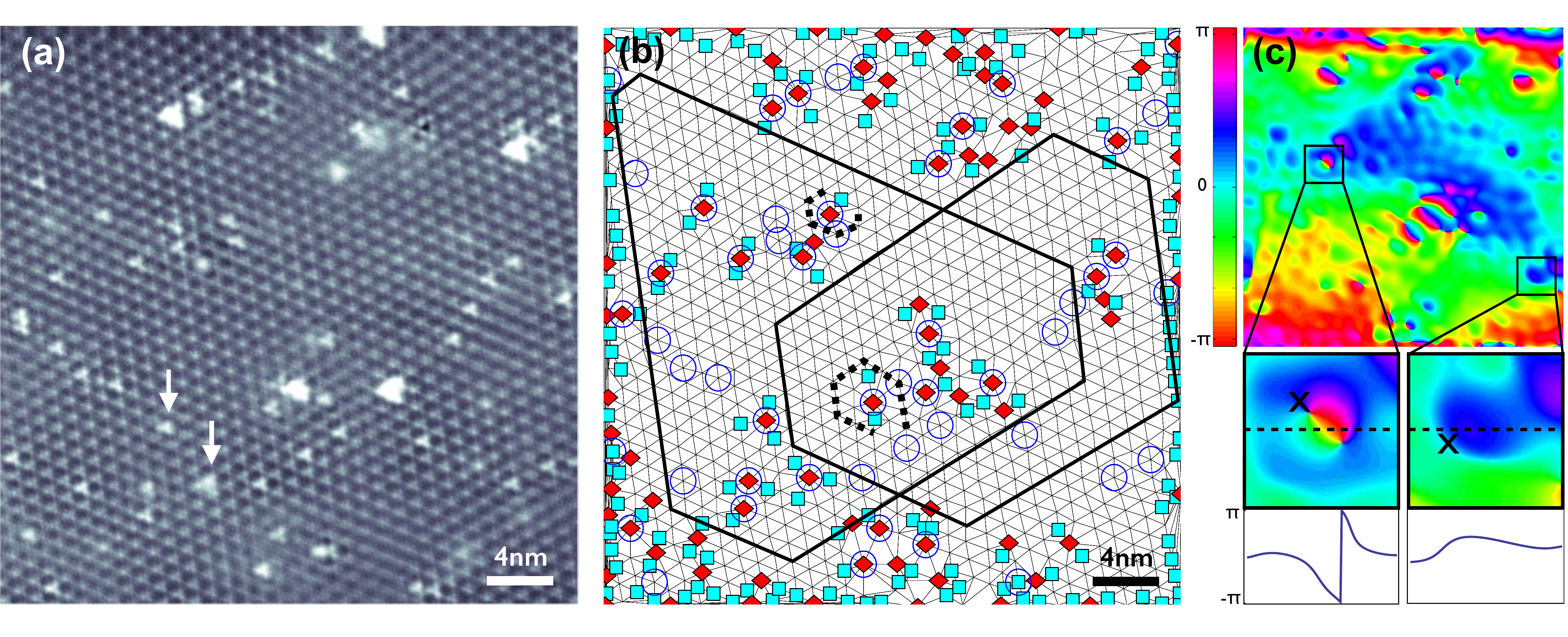}
\caption{(Color online) (a) A topographic image of a $\sim$ 32 nm $\times$ 32 nm region of NbSe$_2$ at $22$ K $<$ $T_c = 34$ K taken under conditions of constant sample-tip current and bias voltage. The heavy white spots are the strong pinning centers. (b) Delaunay analysis of the image. Diamonds (squares) represent CDW maxima with more (fewer) than six edges. Impurity locations are indicated by the circles. Thick broken lines which do not close indicate that some impurities create dislocations.  Larger sized thick solid loops indicate that on larger length scales  there are no free dislocations.  (c) Main panel is the phase configuration of the phase $\phi_1$ of one of the CDW components. The right inset shows a typical smooth modulation near an impurity. The left inset shows a typical vortex-antivortex pair near an impurity. X's are the locations of impurities. The lower panels of the insets are the phase profile along the dotted line.}
\label{topo}
\end{figure*}

We use the scanning tunneling microscopy (STM) data shown in  Fig.~\ref{topo} to obtain real-space information about the spatial dependence of the amplitude $\eta(x)$ and phase $\phi(x)$. The sample used here is the one described in Ref.~\onlinecite{Carlos2013}, and is made by vapor transport. The cleaved surface is believed to be a Se layer, since a Se-Se bond is van deer Waals, while a Se-Nb bond is Coulombic.  Fig.~\ref{topo} (a) shows the STS topographic image of the cleaved surface at $22$ K $<$ $T_c = 34$ K. The voltage and current are fixed to be $-100$ mV and $20$ pA respectively. The measured signal is the vertical displacement of the STM tip; this depends on the physical  topography and on the near Fermi-level electronic density of states at the tip position. The large number of lighter white spots form an approximately triangular lattice with mean lattice constant $\lambda \sim 1$ nm about three times the basic lattice constant, consistent with the CDW wave vector found in scattering measurements  \cite{Moncton1977}. We therefore believe that these are local maxima in $\delta \rho$ arising from CDW formation. The small number of heavy white spots indicate impurities. There are about 40 impurities in this field of view, which contains $\sim 10^3$ CDW unit cells; in other words, the impurity density $n_\text{imp} \simeq 0.4\%$. The signal associated with impurities may come either from a physical change in surface height (associated e.g. with an impurity in the Se layer) or from a change in the local density of states.  However, one may see that in almost all cases the impurity sits in the center of a hexagon of CDW maxima and has a triangular shape of size $\lesssim 1$ nm consistent with interference of three CDW wave vectors. This suggests that a significant contribution of the impurity signal arises from impurity-induced modulations of the density of states, and that in particular impurities lead to an increase in the local density of states which acts as a strong pinning center fixing the local  CDW maximum to the impurity site. More detailed discussions about pinning are given in the supplementary material.



\begin{figure}[tb]
\centering
\includegraphics[width=\columnwidth]{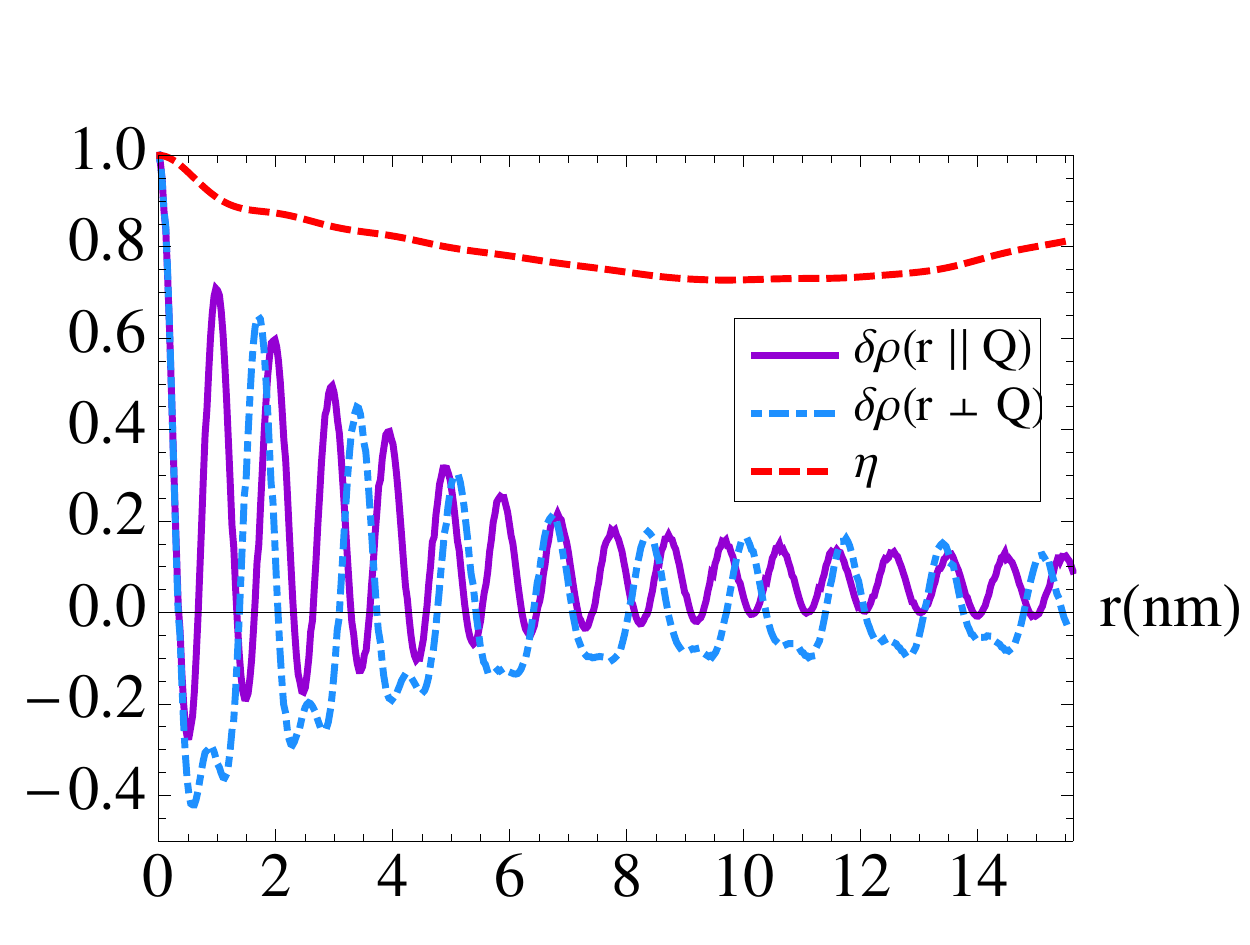}
\caption{(Color online) Autocorrelations of the CDW component $\delta \rho$ parallel to and perpendicular to a CDW wave vector $\vec{Q}_1$, and of the amplitudes $\eta$.}
\label{auto_cdw}
\end{figure}

To analyze this phenomenon more quantitatively, we present, in Fig.~\ref{auto_cdw}, the autocorrelation of the experimental signal, interpreted as a density of states modulation.  We present both the density modulation relative to the average value, $\delta \rho$ and the absolute value or amplitude $\eta$. (The supplementary material explains how the correlation functions are defined and computed).  The amplitude autocorrelation is characterized by an initial rapid decay followed by a more gradual relaxation to a nonzero value while the autocorrelation of the total CDW modulation  $\delta \rho$ decays exponentially with a decay length   $\sim 4$ nm comparable to the inter-impurity spacing $l \approx 5$ nm. Taken together, these facts indicate that impurities correspond to strong pinning centers, but that the main effect of the impurity is on the phase of the CDW order parameter.

While all impurities produce a local maximum in the amplitude of the order parameter, different impurities have different consequences for the phase, shown in Fig.~\ref{topo}(c). The main panel shows the phase field corresponding to one component of the CDW (for details on how this was constructed see the supplementary material). The two insets show expanded views of the phase near impurity sites. The right inset shows an impurity that induces a smooth and small phase modulation. The left inset shows that a different impurity induces a large phase modulation from $-\pi$ to $\pi$ as we move in a counterclockwise fashion around the defect.  Only about $\sim 20\%$ of the identifiable defects produce $2\pi$ phase modulations; the remainder produce smoothly varying modulations of the phase.

Interestingly, at slightly larger distances from the impurity shown in the lower right inset of Fig.~\ref{topo}(c), the phase variation becomes smooth; the impurity actually induces a bound dislocation-antidislocation pair.   Fig.~\ref{topo}(b) presents a Delauney diagram \cite{Murray1990, Grier1991, Dai1992, Dai1993} constructed from the CDW maxima in Fig.~\ref{topo}(a) showing that this is general. Dislocations appear around some impurity sites as shown by the failure of some Delauney loops (shown as broken lines)  to close. However, dislocations are only visible on short length scales; in general loops of size larger than a few lattice constants (solid lines) close, indicating that in this field of view the dislocations appear only in bound dislocation-antidislocation pairs. The loops continue to close even if the size of the loop becomes  as large as the image size, indicating that on the length scales accessible to this experiment, there are no free dislocations.  Therefore, the system is in a Bragg glass phase \cite{Feigel'man1989, Nattermann1990, Bouchaud1991, Bouchaud1992, Korshunov1993, Giamarchi1994, Giamarchi1995, Giamarchi1997, Rosso2004}, with the decay of the $\delta \rho$ autocorrelation being produced primarily by a smoothly varying phase field, as shown in the main panel of Fig.~\ref{topo}(c).

Thus, in summary, the data presented here indicate that the impurities observed in this NbSe$_2$ sample are strong pinning centers, but nevertheless leave the system in a Bragg glass phase, in apparent disagreement with the conventional idea that the strong impurities induce free topological defects and completely destroy the order \cite{Fukuyama1978, Lee1979}. To understand this issue we present an energy analysis inspired by Refs.~\onlinecite{Abe1985} and \onlinecite{Abe1986} and based on the assumption that the impurities are dilute (i.e. the inter-impurity spacing is large compared to the CDW wavelength), but strong. 

A crucial issue in the  analysis is the dimensionality of the system. While NbSe$_2$ has very anisotropic electronic properties \cite{Dordevic2001}, we believe that the appropriate model is three dimensional for the following reasons. First, three dimensional critical scattering is observed  in the similar compound $2H-$TaSe$_2$ \cite{Moncton1977}, with correlation lengths in the in-plane and out-of-plane directions differing only by a factor of three. Second, below the transition temperature, the development of the order parameter agrees with mean-field theory \cite{Moncton1977}, while a two dimensional incommensurate CDW cannot show a true long-range order \cite{Mermin1966}. Third, a first principle calculation showed that single layer NbSe$_2$ does not exhibit the $3\times3$ periodicity \cite{Calandra2009}. These arguments suggest that, most likely because of lattice effect,  the CDW in NbSe$_2$ is not unusually anisotropic. We therefore study a three dimensional model.

For simplicity, we consider a CDW described  by one phase variable $\phi$, and neglect amplitude modulation. We now add impurities at positions $x_a$; these impurities act to locally pin the phase to the values $\theta_a$. At distances $|\vec{x}-\vec{x}_a| \gg \xi$ ($\xi$ is the coherence length of the CDW) the phase will change; this may take place either by a smooth modulation [as shown in the right inset of Fig.~\ref{topo}(c)] or by creation of a defect-antidefect pair [as shown in the left inset of Fig.~\ref{topo}(c)]. 
In the absence of defects the free energy of this phase only model is
\begin{equation}
F =   \int d^3\vec{x}\rho_S  \left( \vec{\nabla} \phi \right)^2 - |V| \sum_{a}  \cos \left[ \theta_a - \phi(\vec{x}_a) \right],
\label{F}
\end{equation}
where $\rho_S$ is the phase stiffness, $x_a$ labels the positions of the impurities, $\theta_a$ is the phase energetically favored by the impurity at $x_a$ (this depends on the position of the impurity), and $V$  is the magnitude of the impurity potential [taken to be the same for all impurities in light of the weak variation of amplitudes found in Fig.~\ref{topo}(a) ]. We have rescaled lengths by the ratio of in-plane to out of plane coherence lengths.  In a simple model, we expect that $\rho_S\sim f_0 |\psi|^2\xi_0^2\sim f_0 t\xi_0^2$ with $f_0$ a measure of the condensation energy per unit volume at $T=0$, $\psi$ the CDW amplitude, $\xi_0$ a bare coherence length, and  $t = (T_c-T)/{T_c}$ the reduced temperature, while $V\sim V_0\psi \sim V_0 \sqrt{t}$ is proportional to a bare pinning potential $V_0$ and to the first power of the CDW amplitude. We assume the impurities are dilute (mean inter-impurity distance $l$ much greater than CDW correlation length $\xi=\xi_0/\sqrt{t}$) \cite{McMillan1978, Weber2011}; this condition breaks down close to the transition temperature, or for dense impurities. 

We now consider the energetics of smoothly varying phase configurations, assuming  for simplicity that $V_0$ is very large. At distances larger than a correlation length from any impurity site, minimization of Eq.~\eqref{F} shows that the phase obeys the Laplace equation $\nabla^2\phi=0$. So a general solution in the three dimensional case is (for $\left|\vec{x} - \vec{x}_a\right|>\xi$)
\begin{equation}
\phi (\vec{x}) = \sum_a \frac{\bar{\theta}_a \xi}{|\vec{x} - \vec{x}_a|},
\label{solutionphase}
\end{equation} 
where $\bar{\theta}_a$'s are parameters to be determined. Substituting this into Eq.~\eqref{F}, we obtain
\begin{equation}
\frac{F}{V} = \frac{\epsilon}{2} \sum_{ab} K_{ab} \bar{\theta}_a \bar{\theta}_b + \frac{1}{2} \sum_a \left(\theta_a - \sum_{b}K_{ab}\bar{\theta}_b \right)^2 .
\label{Fofbartheta}
\end{equation}
where the variable inside the parenthesis is taken to be in the range $[ -\pi, \pi] $,  $\epsilon=8\pi\rho_S/V$, and $I$ is the identity matrix. The kernel is 
$K_{ab}=  \delta_{ab} +\left(1-\delta_{ab}\right) \xi/|\vec{x}_a - \vec{x}_b|$.
Minimizing Eq.~\eqref{Fofbartheta} gives
\begin{equation}
\bar{\theta}_a = \sum_b \left(\epsilon I + K \right)^{-1}_{ab} \theta_b  ,
\label{barthetaeq}
\end{equation}
 
The Coulombic form of  $K$ means that  the inverse matrix $(\epsilon I + K)^{-1}$ has a screening form with a characteristic length $r_\text{TF}=\sqrt{l^3 (1+\epsilon)/4\pi \xi}$; its Fourier components are
\begin{equation}
(\epsilon I + K )^{-1} (p) = \frac{p^2 }{(1+\epsilon) \left( p^2   +r_\text{TF}^{-2} \right)},
\end{equation}
Thus, even if the phases $\theta_a$ preferred by the impurities are random variables, on scales longer than $r_\text{TF}$ fluctuations of the $\bar{\theta}$ are suppressed.   As a result, the variance $\langle \phi (0)^2 \rangle$ is not infra-red divergent and the solution given in Eq.~\eqref{barthetaeq} therefore may have a long ranged order. A further analysis, to be presented in detail elsewhere, shows that  the energy cost of a phase variation of momentum $p<r_{TF}^{-1}$ $\sim p^4$, suggesting both a large number of low energy metastable configurations and that on long scales topological defects may proliferate.

\begin{figure}[tb]
\centering
\includegraphics[width=\columnwidth]{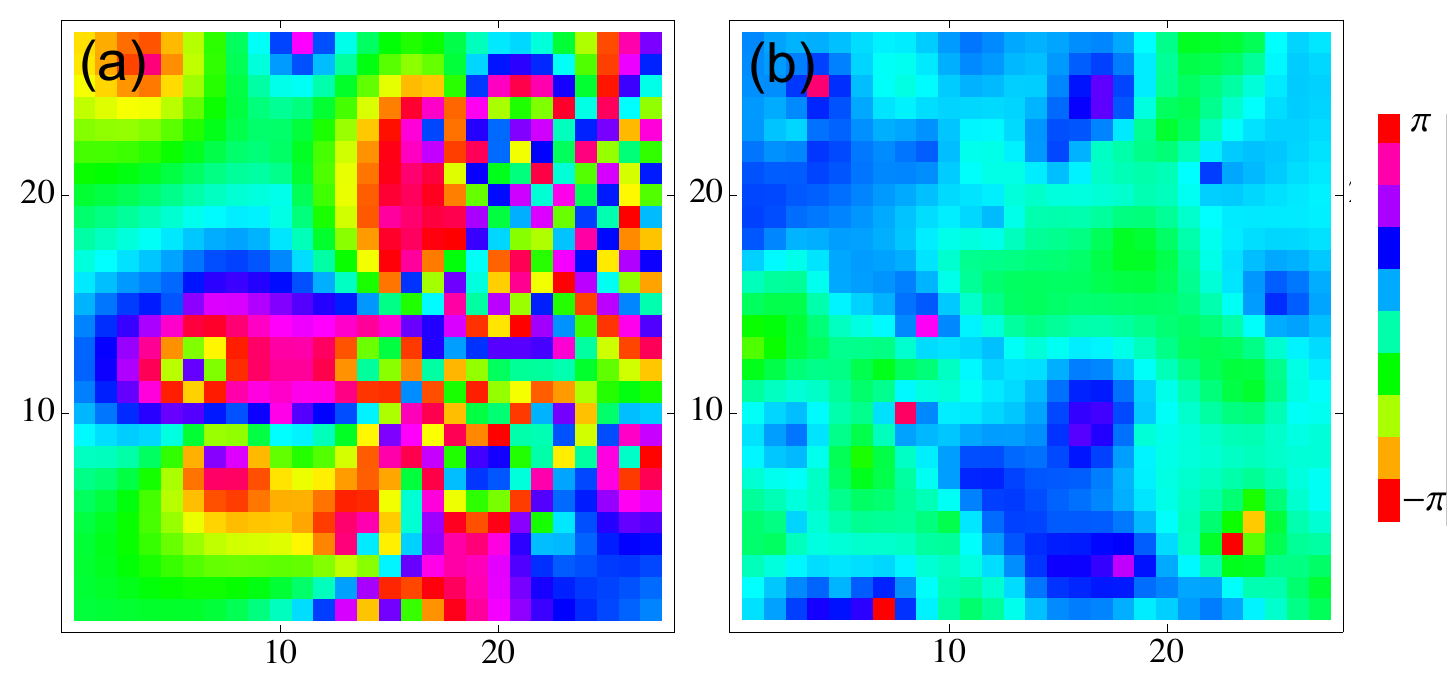}
\caption{(Color online) Typical phase configurations obtained from Eq~\eqref{barthetaeq} when $\epsilon = 0$ in (a) two dimensions, and (b) three dimensions. The lattice constant is $\xi$, and the linear system size is $27\xi$. The impurity concentration is $n_\text{imp} \simeq 0.4\%$.}
\label{num_sol}
\end{figure}

We also numerically solved Eq.~\eqref{barthetaeq} when $\epsilon = 0$ on a regular lattice with a lattice constant $\xi$ and $n_\text{imp} \simeq 0.4\%$. Fig.~\ref{num_sol} shows typical phase configurations in two and three dimensions on a plane (details are presented in the supplementary material). The three dimensional model produces results in qualitative agreement with the data, while the results of the two dimensional model appear to be quite different, in particular producing structure on the scale of $\xi$ rather than the smooth variation actually observed.

We now turn to the question of local topological defects. Making a defect on one site $a$ allows the phase to relax rapidly from the value preferred by the local impurity towards a background value determined by the other defects, decreasing the elastic free energy at the cost of  driving the amplitude to zero over a correlation volume. We may estimate that the defect costs an energy $E_\text{vortex}\sim f_0 t^2 \xi^3 /2 \sim f_0 \sqrt{t} \xi_0^3/2\sim\rho_S\xi$ with $f_0$ the zero temperature condensation energy density defined above. The energy gain is associated with removing one defect from the elastic energy. Using the screened Coulombic form of $K^{-1}$ and noting that the $\theta_a$ are random variables we obtain that   the elastic energy gain is roughly
\begin{equation}
E_\text{elastic} \simeq 4\pi\rho_S\xi (1+\epsilon)^{-1} \theta_a^2 + \mathcal{O} (\xi /l).
\end{equation}
Thus the energy cost of making a defect-antidefect pair is parametrically equal to the cost of the phase deformation and which one is preferred is  determined by an intrinsic property of the CDW [namely the ratio $\kappa =  4\pi \rho_S \xi /E_\text{vortex} (1+\epsilon) $] and the square magnitude of the phase deviation caused by the impurity.  Our finding that about 20$\%$ of impurities induce defects suggests that $\kappa \approx 0.16$, and that defects are only produced when the phase deviates by an amount near its maximal value ($\theta_a \approx \pi$); in analogy with superconductors, the system should be thought of as ``type I'' rather than ``type II".

In summary, we have investigated impurity-induced pinning in the  CDW state of NbSe$_2$, a paradigm charge density wave state. We find  that the impurities are ``strong'' (enhancing the local CDW amplitude by a factor of two), but both experimental and theoretical analyses show that these impurities  lead to a charge density wave phase which varies smoothly over scales parametrically longer than the inter-impurity distance. Only a small fraction of the impurities produce topological defects and these are found to occur only in tightly bound dislocation-antidislocation pairs near the impurities; the material is identified as being in the Bragg glass phase on the scales attainable in the experiment analyzed in this paper. A model analysis shows that  dilute but strong impurities give a long-range order, while the energy cost of phase modulation or creating vortices at large length scales is unusually small. Therefore, in a real system, such a low energy excitation along with many metastable states may lead to a glass phase. The behavior at longer scales is an interesting open problem. 

\begin{acknowledgements}
We thank Rafael M. Fernandes for helpful discussions. This work was supported by Department of Energy Contract Nos. DE-FG02-04ER46157 (J.O.), and DE-FG02-04ER46169 (A.J.M.). STM experiments were supported by the National Science Foundation (NSF) Materials Interdisciplinary Research Team grant number DMR-1122594. Salary support was also provided by the NSF CAREER program under grant DMR-1056527 (E.P.R., A.N.P.). 
\end{acknowledgements}

\end{document}


\title{Supplemental material: Experimental evidence for a Bragg glass density wave phase  in a transition-metal dichalcogenide}

\author{Jun-ichi Okamoto}
\email{okamoto@phys.columbia.edu}
\affiliation{Department of Physics, Columbia University, 538 West 120th Street, New York, New York 10027, USA}

\author{Carlos J. Arguello}
\affiliation{Department of Physics, Columbia University, 538 West 120th Street, New York, New York 10027, USA}

\author{Ethan P. Rosenthal}
\affiliation{Department of Physics, Columbia University, 538 West 120th Street, New York, New York 10027, USA}

\author{Abhay N. Pasupathy}
\affiliation{Department of Physics, Columbia University, 538 West 120th Street, New York, New York 10027, USA}

\author{Andrew J. Millis}
\affiliation{Department of Physics, Columbia University, 538 West 120th Street, New York, New York 10027, USA}

\date{\today}

\maketitle
\section{Bias dependence of topography}
\begin{figure*}[!tb]
  \begin{center}
   \includegraphics[scale= 0.5]{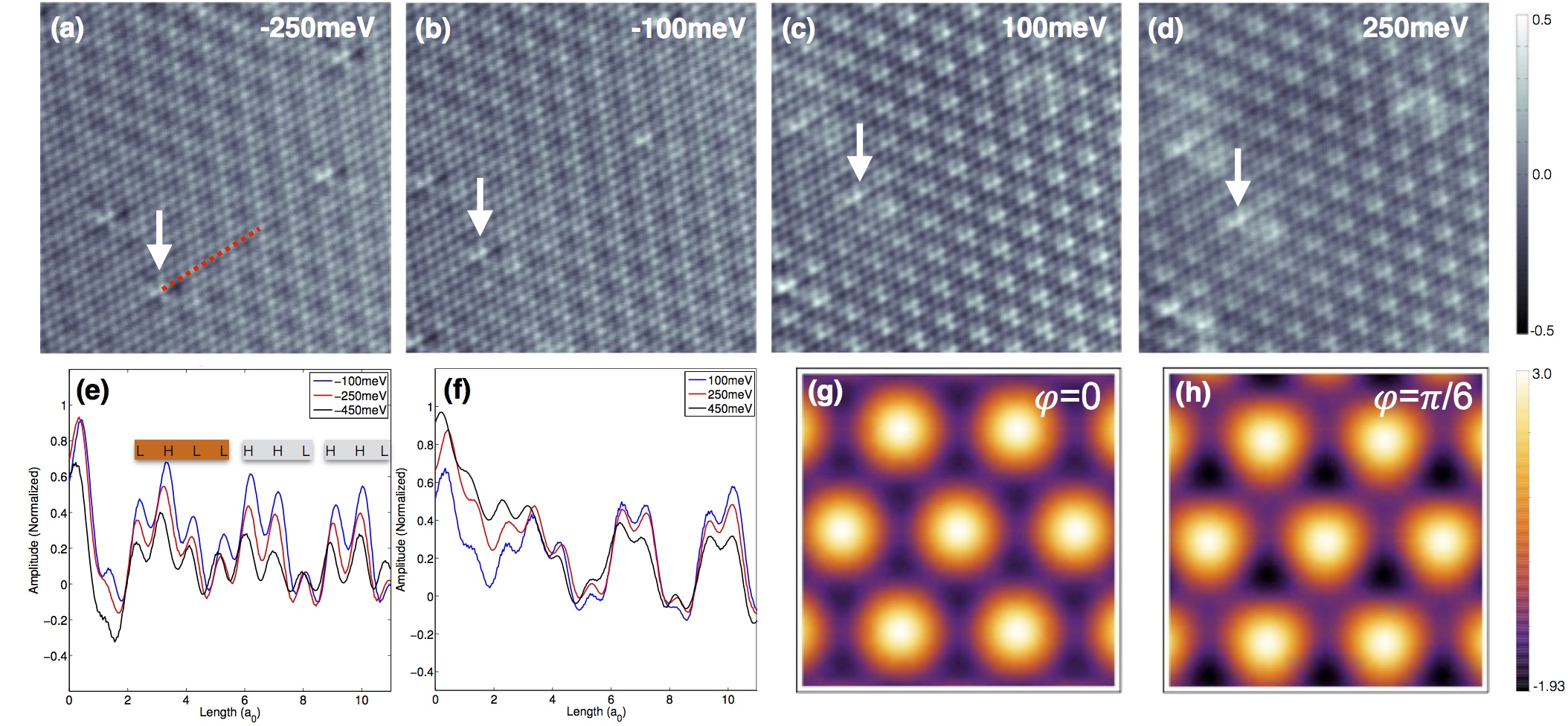}
  \end{center}
  \caption{(a)-(d): topographic pictures at different bias voltages. Arrows indicate the same impurity site at difference voltages. The intensity along the dotted line, which includes an impurity at the left end, is given in (e) and (f) for negative and positive voltages respectively. In these, we set the absolute maxima of the intensity in each picture to 1. In (e), the periodicity of CDW is expressed by the ordering of high maximum ``H", and low maximum ``L". Near the impurity, the periodicity is distorted from the ideal one  ``HHL". (g) and (h) are the density modulations at $\varphi = 0$ and $\varphi  = \pi/6$ with $\eta = 1$ in Eq.~\eqref{drho}. }
  \label{bias}
\end{figure*}

In Figs.~\ref{bias}(a)-(d), we show the bias voltage dependence of topographic pictures, whose amplitudes are normalized such that maximum and minimum are fit into $[-0.5, 0.5]$. The following analysis of these pictures shows (i) the intensity modulations around impurities are mainly due to local density modulations, and (ii) impurities act as strong pinning centers.

First, as the voltage increases from the negative one to the positive one, we see that the shape of CDW changes from a configuration with three deep minimum surrounding one maxima to the one with six shallow minimum. This change in the number of minimum is due to the change of CDW phases, and can be understood as follows. The total CDW density is given by the linear superposition of three CDWs as
\begin{equation}
\delta \rho (\vec{x}) = \eta \sum_{i=1,2,3} \cos (\vec{Q_i} \cdot \vec{x} + \phi_i ),
\label{drho}
\end{equation}
with the phases of three CDWs $\phi_i$, the CDW amplitude $\eta$, and the ordering vectors $\vec{Q}_i$. The three phases $\phi_i$ can be decomposed into the relative displacement vector $\vec{u}$ and the total phase $\varphi$,
\begin{equation}
\phi_i = \vec{u} \cdot \vec{Q_i} + \varphi.
\end{equation}
We depicted the density modulation given by Eq.~\eqref{drho} with $\varphi  = 0$ and $\pi/6$ in Figs.~\ref{bias}(e) and \ref{bias}(f). With $\varphi = 0$, each maxima is surrounded by six equivalent shallow minimum like a Kagome lattice, while the degeneracy among six is lifted once we make $\varphi \neq 0$. At $\varphi = \pi/6$, there are three deep minimum surrounding one maxima. Therefore, at the negative voltages $\varphi $ is close to $\pi/6$, and at positive voltages $\varphi$ seems closer to $0$. The intensity modulations around impurities such as the one near the arrows in Fig.~\ref{bias} follow the same change as the voltage changes, indicating that these modulations are triggered by the same mechanism as other CDW modulations, i.e., local density modulations not the height modulations. 

Second, we also see in Fig.~\ref{bias}(a)-(d) that impurity sites are always at CDW maximum, and that the CDW amplitudes near them are enhanced, as more clearly presented in the line-cut pictures in Figs.~\ref{bias}(e) and \ref{bias}(f). For instance, in Fig.~\ref{bias}(e), in the vicinity of the impurity, the ideal periodicity, two maximum followed by one minima, is distorted to have a maxima at the impurity, and the CDW amplitude is increased as well. These observations indicate that impurities strongly pin the CDWs and distort them locally.

\section{Defining order parameters}
\begin{figure*}[!tb]
  \begin{center}
   \includegraphics[scale= 0.4]{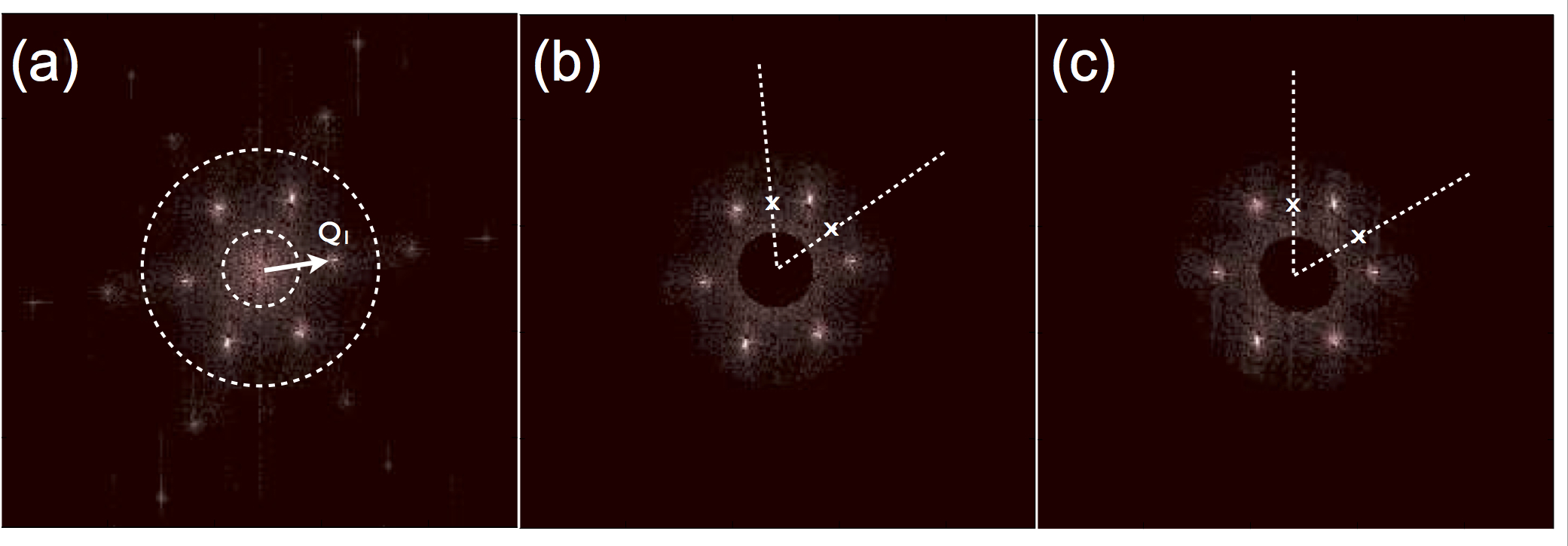}
  \end{center}
  \caption{(a) The absolute values of Fourier components $\delta \rho (\vec{k})$. The Fourier components outside the annulus given by the dashed lines are removed. (b) Fourier components after removing atomic peaks and $\vec{k} =0$ components. X's are the middle points of neighboring CDW peaks.  (c) The Fourier components after affine transformations.}
  \label{affine_fou}
\end{figure*}

\begin{figure*}[!tb]
  \begin{center}
   \includegraphics[scale= 0.4]{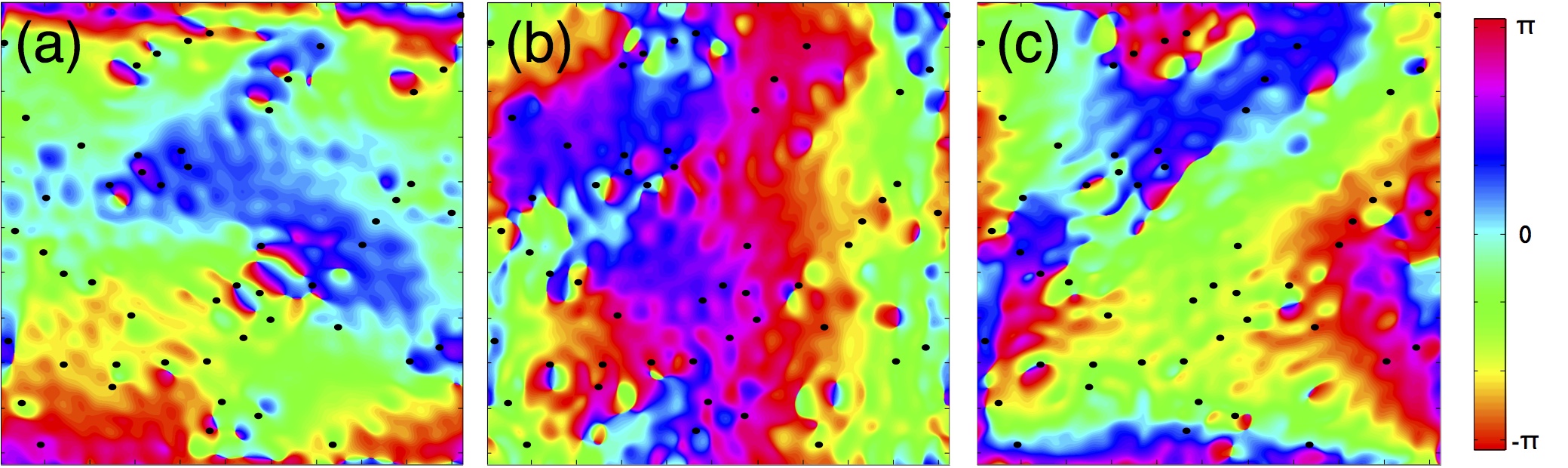}
  \end{center}
  \caption{The phases of three order parameters. Black dots represent the locations of impurities. The gradual change along the horizontal axis in (b) is an artifact in order to take the ordering vector $\vec{Q}_i$ such that $\sum_i \vec{Q}_i =0$; $\vec{Q}$ are specified only through integer numbers due to the periodic boundary conditions for the Fourier transformation.  }
  \label{phases}
\end{figure*}

In this section we explain how we process the data and define the order parameters. The first step is to  extract the CDW component from the topographic image shown in Fig.~1 in the main text. To do this, we Fourier transformed the data (an image of $1024\times1024$ pixels) assuming that the field of view is periodically repeated in the $x$ and $y$ directions. The Fourier transformed image has a central peak, atomic Bragg peaks at reciprocal lattice vectors $\vec{G}_i$, and $6$ relatively sharp peaks originating from the CDW correlations located approximately on the vertices of a hexagon, $\vec{Q}_{i=1...6}$ [Fig.~\ref{affine_fou}(a)]. We define the mean distance of the CDW peaks from the origin to be $Q\approx |\vec{G}|/3$. We then filter the data by retaining only the region in an annulus of inner radius $Q/2$ and outer radius $3Q/2$ [Fig.~\ref{affine_fou}(b)]. Next, we performed an affine transformation
\begin{equation}
\vec{x}' =  A \vec{x} + \vec{b}
\label{affine}
\end{equation} 
to remove the shear distortion, which we believe arises from the drift of the scanning tunneling microprobe. To define the affine transformation we partition the annulus into six segments, with each boundary between segments passing through a middle point of neighboring CDW peaks and through the origin. In Fig.~\ref{affine_fou}(b), X's are the middle points, and dashed lines are the boundaries of a segment. For each segment, we define an affine transformation which maps each middle point position to a vertex of a regular hexagon. The positions of two middle points and the origin of the picture uniquely defines the parameters, $A$ and $\vec{b}$, in Eq.~\eqref{affine} for each segment. After the affine transformations, the Fourier components will look like Fig.~\ref{affine_fou} (c). Finally, we need to perform the inverse Fourier transform of the filtered, affine-transformed data to obtain a position dependent amplitude which we identify with the CDW component $\delta \rho (\vec{x})$ in the real space. $\delta \rho (\vec{x})$ is related to the order parameters, $\psi_i (\vec{x}) = \eta_i(\vec{x}) e^{i \phi_i (\vec{x})}$ by 
\begin{equation}
\delta \rho (\vec{x}) = \sum_{i=1}^3 \Re \psi_i (\vec{x}) e^{ i \vec{Q}_i \cdot \vec{x}}.
\end{equation}
To obtain the $\eta_i$ and $\phi_i$, we shift each CDW peak in a circle of radius $Q/2$ to the origin of the Fourier space, and Fourier transform it to the real space. The obtained phases of the order parameters are given in Fig.~\ref{phases}.

\section{Delaunay diagram}
\begin{figure*}[!tb]
  \begin{center}
   \includegraphics[scale= 0.4]{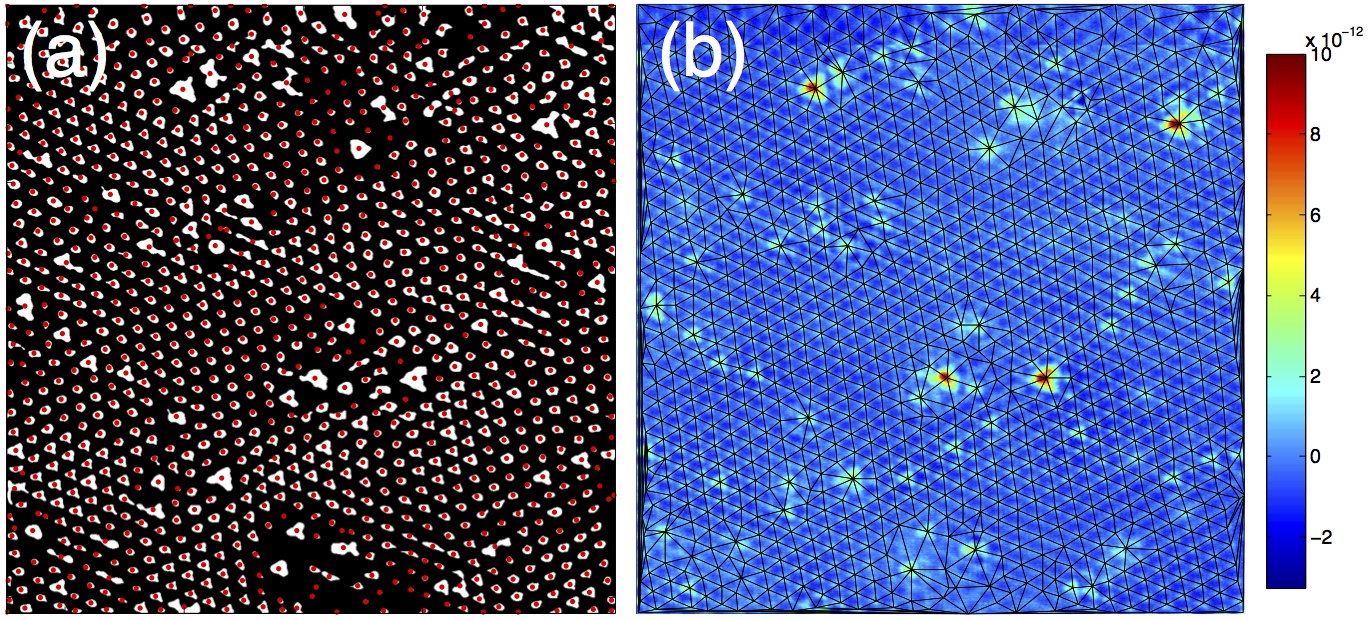}
  \end{center}
  \caption{(a) The binary image after filtering the topographic image by a threshold value. (b) The obtained Delaunay image overlaid on the topographic image.}
  \label{bw}
\end{figure*}

In this section, the detailed procedures of obtaining the Delaunay diagram is presented. First, we remove the central peaks of the Fourier image as Fig.~\ref{affine_fou}, and then transform it back to the real space. Then, we create a binary image by distinguishing each data points by a threshold value; in Fig.~\ref{bw}, the white parts are above the value, and the black parts are below. The threshold value is chosen such that we obtain a maximal number of distinct points. If the value is too low neighboring maximal islands are connected, while if the value is too large, we miss some of the CDW maxima. The locations of maxima are identified by red points in Fig.~\ref{bw}. The Delaunay diagram is then constructed based on these points.

\section{Autocorrelations}
In this section, we explain how we computed the autocorrelations of the charge-density wave (CDW) components $\delta \rho (\vec{x})$, of the amplitude $\eta(\vec{x})$, and of the phase $\phi (\vec{x})$.

An unbiased autocorrelation of a quantity $A(\vec{x})$ confined in a finite size of $L \times L$ is defined as
\beq
\langle A(\vec{x}) A(0)\rangle  = \frac{\int d^2 y A\left( \vec{x}+\vec{y} \right) A\left( \vec{y} \right) }{\int d^2 y A\left(\vec{y} \right) A\left( \vec{y} \right)} \times \frac{L^2}{\text{Overlapped area}}. 
\label{auto}
\eeq
The last factor normalizes the autocorrelation depending on the shifting vector $\vec{x}$. The integration becomes a summation for a discrete quantity.

For the CDW component, we angular averaged the autocorrelations about $\pi/3$ rotations;
\begin{equation}
\bar{A}(\vec{x}) \equiv \frac{1}{3} \left[ A(\vec{x}) +  A(\vec{x}') +  A(\vec{x}'')  \right],
\end{equation}
where $\vec{x}'$, and $\vec{x}''$ are the rotated coordinates of $\vec{x}$ by $\pi/3$ and $2\pi/3$.

\section{Numerical solutions}
In this section we present the details of the numerical analysis in the strong pinning limit, assuming also that the CDW correlation length $\xi$ is small compared to the mean inter-impurity distance $l$.
Since the phase of the CDW $\phi(\vec{x})$ obeys the Laplace equation away from impurities $( \nabla^2 \phi = 0 )$, we use the following ansatz, valid for positions $x$ farther than $\xi$ from any impurity, i.e.  $\left|\vec{x} - \vec{x}_a \right|>\xi$,
\begin{equation}
\phi(\vec{x}) = \sum_a \frac{\bar{\theta}_a \xi}{\left|\vec{x} - \vec{x}_a \right|} ,
\label{ansatz}
\end{equation}
while if $\left|\vec{x} - \vec{x}_a \right|\leq\xi$ for some $a$ then
\begin{equation}
\phi(\vec{x}) = \bar{\theta}_a+ \sum_{b\neq a} \frac{\bar{\theta}_b \xi}{\left|\vec{x}_a - \vec{x}_b \right|} .
\label{ansatz1}
\end{equation}
Here the $\bar{\theta}$ are parameters characterizing the fundamental solution. An impurity at site $a$ acts to pin the phase to a value $\theta_a$ (determined by the position of the impurity relative to the lattice sites). In the strong pinning limit, the phase at each impurity site is nearly equal to the phase preferred by the impurity, so we approximate the cosine potential by a quadratic one. Substituting the ansatz to the energy functional, we find
\begin{equation}
\frac{F}{V} = \frac{\epsilon}{2} \sum_{ab} K_{ab} \bar{\theta}_a \bar{\theta}_b + \frac{1}{2} \sum_a \left(\theta_a - \sum_{b}K_{ab}\bar{\theta}_b \right)^2,
\end{equation}
with a kernel $K$ given by
\begin{equation}
K_{ab} =  \delta_{ab} +\left(1-\delta_{ab}\right)\frac{ \xi}{|\vec{x}_a - \vec{x}_b|}.
\label{kernel}
\end{equation}
Minimizing the free energy by $\bar{\theta}_a$ leads to
\begin{equation}
\bar{\theta}_a = \sum_{b} (\epsilon I  + K )^{-1}_{ab} \theta_b,
\label{bartheta}
\end{equation}
where $I$ is the identity matrix.

Now we explain the procedures how to obtain numerical solutions of Eq.~\eqref{bartheta} when $\epsilon = 0$. A regular lattice with a lattice constant $\xi$ is used, and impurities are randomly put on the lattice with phases randomly chosen from $-\pi$ to $\pi$. Then, the matrix $K$ in Eq.~\eqref{kernel} is constructed. In two dimensions, the solution of the Laplace equation becomes logarithmic, so $K_{ab}$ should be replaced to
\begin{equation}
K_{ab} = \delta_{ab} + (1-\delta_{ab}) \log\left( \frac{|\vec{x}_a - \vec{x}_b|}{\xi}\right).
\end{equation}
The impurity densities are both set to $n_\text{imp} \approx 0.4\%$; the average impurity distances are $l \approx 5\xi$ and $2.9\xi$ in two and three dimensions respectively. The phase $\phi$ on each lattice site is computed from Eqs.~\eqref{ansatz}, \eqref{ansatz1}, and \eqref{bartheta}. The size of the images, $L = 27\xi$, is comparable to the topographic image, $L \approx 32\xi$. Quantitatively, the two dimensional case shows a much more rapid oscillations compared to the three dimensional case; such a rapidly varying structure is not observed in the topographic image.  

